\documentclass{article}

\usepackage[utf8]{inputenc}
\usepackage[top=1.25in, bottom=1.25in, left=1.25in, right=1.25in]{geometry}
\usepackage{graphicx}
\usepackage{subfigure}
\usepackage{amsfonts}
\usepackage{amsgen}
\usepackage{amsbsy}
\usepackage{amsmath}
\usepackage{amssymb}
\usepackage{mathrsfs}
\usepackage{mathtools}
\usepackage{epstopdf}
\usepackage{caption}
\usepackage{tabularx}
\usepackage{tabu}
\usepackage{xcolor}
\usepackage{nicefrac}
\usepackage{enumitem}
\usepackage{bm}
\usepackage{natbib}
\usepackage{url}
\usepackage{bbold}
\usepackage{grffile}
\usepackage{breqn}
\usepackage[bbgreekl]{mathbbol}
\usepackage{bbm}

\usepackage{chngcntr} 

\newcommand\ie{\textit{i.e. }}

\newcommand\eg{\textit{e.g. }}

\newcommand\etc{\textit{etc}}
\DeclarePairedDelimiter\abs{\lvert}{\rvert}

\title{Implementation note on a minimal hybrid lubrication/granular dynamics model for dense suspensions}
\author{Zhouyang Ge%
  \footnote{Corresponding author. Email: \texttt{zhoge@mech.kth.se}. Last modified: \today.}
  , Luca Brandt}
\date{\small \it Linn\'e Flow Centre and SeRC (Swedish e-Science Research Centre), Department of Engineering Mechanics,\\
  KTH Royal Institute of Technology, S-100 44 Stockholm, Sweden}

\begin{document}

\maketitle

\begin{abstract}
  We describe and summarize a class of minimal numerical models emerged from recent development of simulation methods for
  dense particle suspensions in overdamped linear flows.
  The main ingredients include
  (i) a frame-invariant, short-range lubrication model for spherical particles, and
  (ii) a soft-core, stick/slide frictional contact model activated when particles overlap.
  We implement a version of the model using a modified velocity-Verlet algorithm that
  explicitly solves the $N$-body dynamical system in $\mathcal{O}(cN)$ operations,
  where $c$ is a kernel constant depending on the cutoff of particle interactions.
  The implementation is validated against literature results on jamming transition and shear thickening suspensions from 40\% to 64\% volume fractions.
  Potential strategies to extend the present methodology to non-spherical particles are also suggested for very concentrated suspensions.
\end{abstract}

\section{Introduction}

\begin{figure}
  \centering
  \includegraphics[width=0.44\columnwidth]{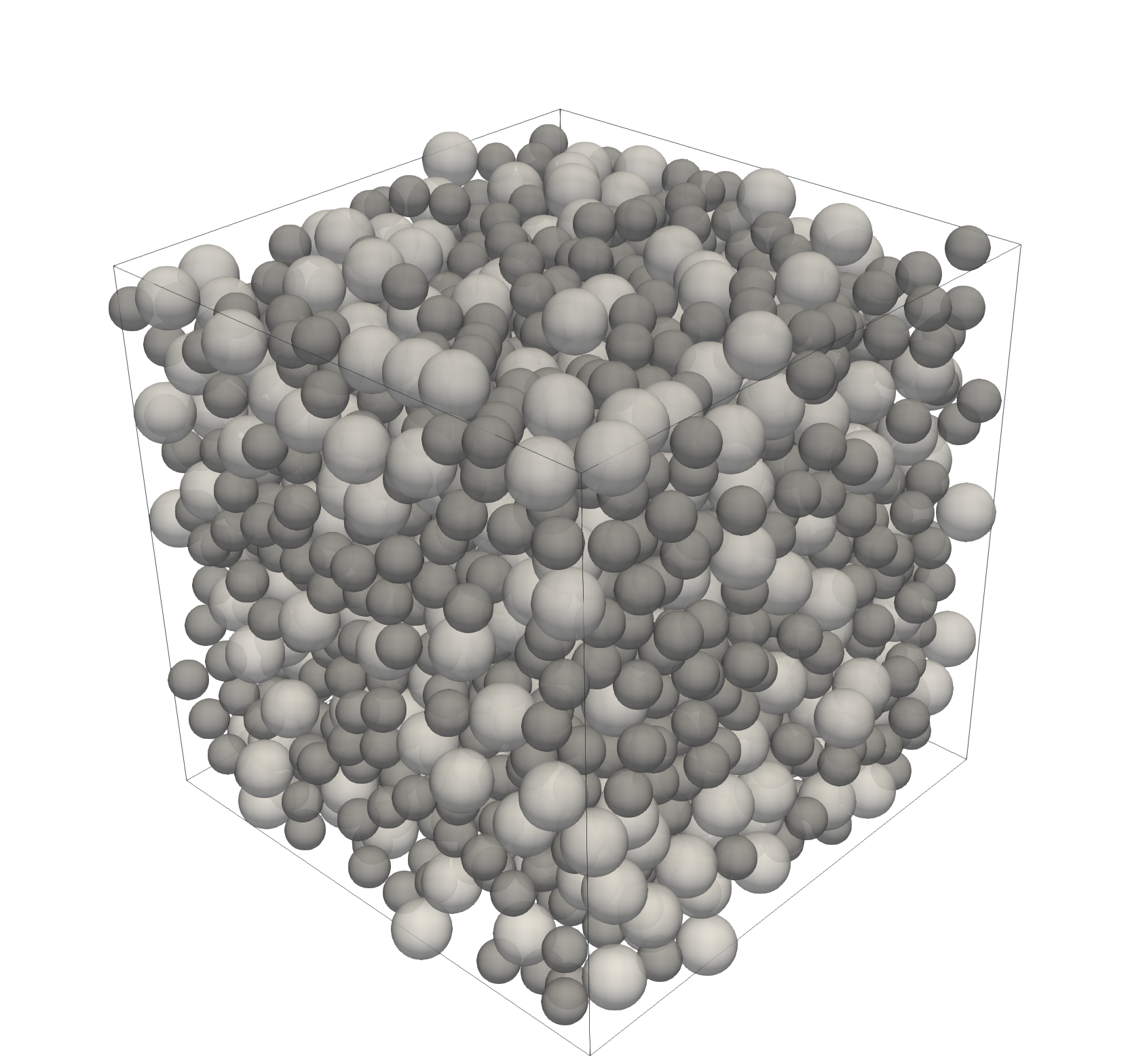}
  \includegraphics[width=0.4\columnwidth]{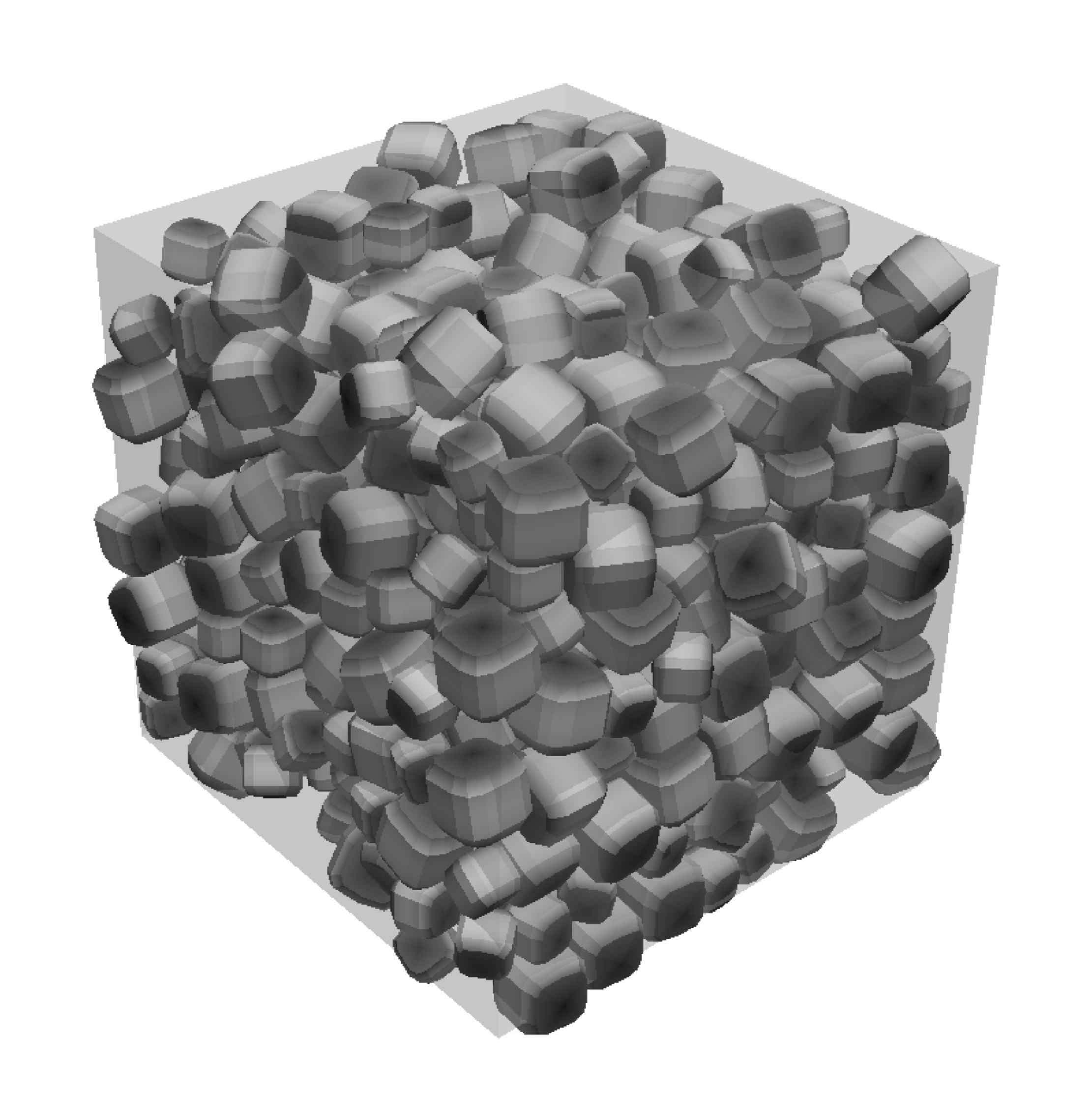}
  \caption{Visualizations of dense particle suspensions. (left) A bidisperse suspension of 2000 spheres at 55\% volume fraction. (right) A random organization of 500 superballs at 50\% volume fraction.}
  \label{fig:snap}
\end{figure}

The behaviour of systems involving the motion of small particles in a suspending fluid covers a wide range of phenomena of interest to both scientists and engineers. Dense suspensions, where the volume fraction of solid particles becomes comparable to or even higher than that of the fluid (see Figure \ref{fig:snap}), have particularly rich and sometimes unexpected rheologies, such as yielding, shear thinning, continuous shear thickening (CST) or discontinuous shear thickening (DST) \citep{mewis_wagner_book, Morton_Morris_2014, guazzelli_pouliquen_2018, Morris_annurev2020}. Apart from being theoretically intriguing, these complex behaviours often have major practical implications. For instance, while it makes sense for the cement industry to manufacture suspensions that do not shear thicken, the same feature becomes an advantage for designing flexible body armor.

Despite the practical importance, theoretical development of suspension rheology remains challenging and
only a few analytical solutions have been found in the dilute regime, see \eg \cite{Einstein_1906, batchelor_green_1972b}.
This is partially due to the lack of a precise knowledge or control of the various interactions at the particle level and
partially due to the mathematical difficulties involved in many-body problems.
On the other hand, solving a system of interacting particles appears relatively straightforward in an algorithmic perspective.
In fact, the last few decades have seen tremendous advancement in both numerical simulations and computer hardware.
In the context of rheology and soft condensed matter, some of the numerical models that have been developed include
molecular dynamics \citep{Alder_Wainwright_1959,Verlet1967},
dissipative particle dynamics \citep{Hoogerbrugge_1992, Groot_Warren_1997},
Stokesian dynamics \citep{Brady_Bossis1988}
and hybrid lubrication/granular dynamics (HLGD)%
\footnote{We dub it this way not because we invented the model but because there is no unique identifier in the literature.}
\citep{Mari_Seto_2014JoR, Cheal_Ness_2018}, to name a few.
The objective of the present note is thus to provide a brief summary of the latest development;
specifically, we focus on HLGD that recently emerged as an effective method to minimally model dense particle suspensions in low-Reynolds-number flow.
It is our hope that the numerical details presented in the following will help beginners to this field gain a clear idea in how to implement the seemingly complex algorithm, which are usually scattered in the supplementary information of papers focusing on rheology or physics.

The paper is organized as follows. In Sec.\ \ref{sec:dlcd}, we summarize the mathematical formulation, relevant physical parameters and their rheological characterisation. A brief note on the numerical integration of the equations of motion is also provided. In Sec.\ \ref{valid}, we validate the implementation against two benchmark cases in the literature, \ie jamming transition and shear thickening. Finally, we end with a short discussion on potential extensions of the present methodology to non-spherical particles.

\section{The HLGD model}
\label{sec:dlcd}

In this section, we briefly summarize the mathematical formulation and essential physical parameters of the hybrid lubrication/granular dynamics. Detailed descriptions can be found in \cite{Seto_PRL2013, Mari_Seto_2014JoR, Cheal_Ness_2018} and references therein.

\subsection{Mathematical formulation}
\label{sec:eqs}

The translational and rotational dynamics of a rigid particle is governed by the Newton-Euler equations, 
\begin{subequations} \label{newton-euler}
  \begin{equation} 
    \begin{aligned} \label{force-balance}
      \sum_M {\bm F}_i^M = m_i \frac{d{\bm u}_i}{dt}, 
    \end{aligned}
  \end{equation}
  \begin{equation} 
    \begin{aligned}
      \sum_M {\bm T}_i^M = {\bm I}_i \frac{d{\bm \omega}_i}{dt} + {\bm \omega}_i\times({\bm I}_i{\bm \omega}_i),
    \end{aligned}
  \end{equation}
\end{subequations}
where ${\bm F}_i$ and ${\bm T}_i$ denote the force and torque exerted on the center-of-mass of particle $i$; $m_i$ and ${\bm I}_i$ are its mass and moment-of-inertia tensor in the body frame (scalar for spheres); ${\bm u}_i$ and ${\bm \omega}_i$ denote its translational and angular velocities, respectively. The force and torque are summed over various modes of particle-fluid and particle-particle interactions that include the following single-body or pairwise \emph{hydrodynamic}, \emph{contact}, and \emph{physico-chemical} contributions
\begin{subequations}
  \begin{equation} 
    \begin{aligned} \label{force-decomp}
      \sum_M {\bm F}_i^M = {\bm F}_i^S + \sum_j^{N_L} {\bm F}_{i,j}^L + \sum_j^{N_C} {\bm F}_{i,j}^C + \sum_j^{N_R} {\bm F}_{i,j}^R + \sum_j^{N_A} {\bm F}_{i,j}^A, 
    \end{aligned}
  \end{equation}
  \begin{equation} 
    \begin{aligned}
      \sum_M {\bm T}_i^M = {\bm T}_i^S + \sum_j^{N_L} {\bm T}_{i,j}^L + \sum_j^{N_C} {\bm T}_{i,j}^C + \sum_j^{N_R} {\bm T}_{i,j}^R + \sum_j^{N_A} {\bm T}_{i,j}^A.
    \end{aligned}
  \end{equation}
\end{subequations}
Their specific functional forms, for the case of spheres, are listed below.

\begin{enumerate}

\item Stokes' drag (acting on each particle $i$)
  \begin{subequations}
    \begin{equation} 
      \begin{aligned}
        {\bm F}^S_i = -6\pi\mu a_i({\bm u}_i - {\bm U}^\infty_i), 
      \end{aligned}
    \end{equation}
    \begin{equation} 
      \begin{aligned}
        {\bm T}^S_i = -8\pi\mu a_i^3({\bm \omega}_i - {\bm \Omega}^\infty_i),
      \end{aligned}
    \end{equation}
  \end{subequations}
  where $\mu$ is the dynamic viscosity of the underlying fluid, $a_i$ the particle radius, ${\bm U}^\infty_i$ and ${\bm \Omega}^\infty_i$ the undisturbed translational and angular velocities of the fluid at the particle position. A linear flow satisfies the relation,
  ${\bm U}^\infty={\bm \Omega}^\infty \times {\bm x} + \mathbb{E}^\infty \cdot {\bm x}$, where $\bm x$ denotes the position vector, and $\mathbb{E}^\infty$ is the rate-of-strain tensor \citep{Batchelor}.
  
\item Lubrication (acting on each lubricating pair $i,j$)
  \begin{subequations}
    \begin{equation} 
      \begin{aligned}
        {\bm F}^L_{i,j} = -(X_{ii}^A \mathbb{P}_n + Y_{ii}^A \mathbb{P}_t)({\bm u}_i-{\bm u}_j) \\
        + Y_{ii}^B ({\bm \omega}_i \times {\bm n}_{ij}) 
        + Y_{ji}^B ({\bm \omega}_j \times {\bm n}_{ij}),
      \end{aligned}
      \label{lub F}
    \end{equation}
    \begin{equation} 
      \begin{aligned}
        {\bm F}^L_{j,i} = -{\bm F}^L_{i,j},
      \end{aligned}
      \label{lub F 2}
    \end{equation}
    \begin{equation} 
      \begin{aligned}
        {\bm T}^L_{i,j} = -Y_{ii}^B({\bm u}_i-{\bm u}_j) \times {\bm n}_{ij} 
        - \mathbb{P}_t(Y_{ii}^C{\bm \omega}_i + Y_{ij}^C{\bm \omega}_j),
      \end{aligned}
      \label{lub T}
    \end{equation}
    \begin{equation} 
      \begin{aligned}
        {\bm T}^L_{j,i} = -Y_{ji}^B({\bm u}_i-{\bm u}_j) \times {\bm n}_{ij} 
        - \mathbb{P}_t(Y_{ji}^C{\bm \omega}_i + Y_{jj}^C{\bm \omega}_j),
      \end{aligned}
      \label{lub T 2}
    \end{equation}
  \end{subequations}
  where ${\bm n}_{ij}$ denotes the unit normal vector pointing from particle $i$ to particle $j$, $\mathbb{P}_n={\bm n}_{ij}{\bm n}_{ij}$ and $\mathbb{P}_t=\mathbb{1}-{\bm n}_{ij}{\bm n}_{ij}$ represent the normal and tangential projection matrices and the $X$'s and $Y$'s are scalar resistances depending on $\mu$, $a_i$, $a_j$, and the gap between the two particles; see Appendix \ref{app: scalar resist} for the detailed expressions and Appendix \ref{app: lub range} for their numerical treatment. 

\item Contact force (acting on each overlapping pair $i,j$)
  \begin{subequations}
    \begin{equation} 
      \begin{aligned}
        {\bm F}^C_{i,j} = -k_n {\bm h}_{ij}  - \gamma_n \mathbb{P}_n({\bm u}_i-{\bm u}_j) - k_t {\bm \xi}_{ij},
      \end{aligned}
      \label{col F}
    \end{equation}
    \begin{equation} 
      \begin{aligned}
        {\bm F}^C_{j,i} = - {\bm F}^C_{i,j},  \quad \textrm{with} \quad 
        \abs{k_t {\bm \xi}_{ij}} \leq 
        \mu_c \abs{k_n {\bm h}_{ij}  + \gamma_n \mathbb{P}_n ({\bm u}_i-{\bm u}_j) },
      \end{aligned}
      \label{col F 2}
    \end{equation}
    \begin{equation} 
      \begin{aligned}
        {\bm T}^C_{i,j} = a_{i} k_t ({\bm n}_{ij} \times {\bm \xi}_{ij}),
      \end{aligned}
    \end{equation}
    \begin{equation} 
      \begin{aligned}
        {\bm T}^C_{j,i} = a_{j} k_t ({\bm n}_{ij} \times {\bm \xi}_{ij}).
      \end{aligned}
    \end{equation}
  \end{subequations}
  where ${\bm h}_{ij}=h_{ij}{\bm n}_{ij}$ denotes the signed normal surface gap between particles $i$ and $j$ (positive when overlapping), ${\bm \xi}_{ij}$ the signed tangential stretch (see Appendix \ref{app: tang stretch} for definition), $k_n$ the normal spring constant, $\gamma_n$ the damping constant, $k_t$ the tangential spring constant, and $\mu_c$ the friction coefficient. The condition in Eq.\ \eqref{col F 2} states the Coulomb's law of friction.

\item Electrostatic repulsion (acting on each repulsing pair $i,j$)
  \begin{subequations}
    \begin{equation} 
      \begin{aligned}
        {\bm F}^R_{i,j} = -F_{er}\bar{a}/a_1 \exp(-\kappa h_{ij}) {\bm n}_{ij}, 
      \end{aligned}
    \end{equation}
    \begin{equation} 
      \begin{aligned}
        {\bm F}^R_{j,i} = -{\bm F}^R_{i,j}.
      \end{aligned}
    \end{equation}
  \end{subequations}
  where $F_{er}$ is the force scale for the electrostatic repulsion, $\bar{a}=2a_ia_j/(a_i+a_j)$ the harmonic mean radius of two interacting particles $i$ and $j$, and $\kappa$ the inverse Debye length.

\item Van der Waals attraction (acting on each attracting pair $i,j$)
  \begin{subequations}
    \begin{equation} 
      \begin{aligned}
        {\bm F}^A_{i,j} = \frac{A\bar{a} {\bm n}_{ij}}{12(h_{ij}^2+\epsilon^2)}, 
      \end{aligned}
    \end{equation}
    \begin{equation} 
      \begin{aligned}
        {\bm F}^A_{j,i} = -{\bm F}^A_{i,j}.
      \end{aligned}
    \end{equation}
  \end{subequations}
  where $A$ is the Hamaker constant and $\epsilon=0.1\bar{a}$ a regularization term \citep{Singh_attr_prl2019}.

\end{enumerate}

Note that the hydrodynamic force and torque given above are strictly valid for spheres. The validity of these expressions for non-spherical particles depends on (i) the dominance of the non-hydrodynamic interactions over the hydrodynamic ones and (ii) the departure of the particle shape from a sphere. In addition, the torque due to non-hydrodynamic forces has the general form of ${\bm T}= {\bm r} \times {\bm F}$, where ${\bm r}$ is the lever arm vector. Therefore, extra care must be taken when applying the above formulation directly to non-spherical particles.

\subsection{Model parameters}
\label{model param}

The preceding equations are formulated in dimensional form with a unit system flexibly chosen for the convenience of simulations (\eg $a_1=1$ [{\small Length}], $\dot{\gamma}=10^{-2}$ [{\small 1/Time}], \etc). For the model output to correspond physically to a dense suspension of inertialess, rigid particles, the following asymptotic conditions must be satisfied as close as possible.

\begin{enumerate}

\item A vanishing Stokes number, St $=\rho \dot{\gamma} a_1^2/\mu \ll 1$.

  The Stokes number controls the effect of particle inertia on the particle dynamics in viscous flows. This can be readily seen by inserting $F \sim \mu a U$ in the force balance $F \approx \rho a^3 (\delta U/ \delta \tau)$, and comparing $\delta \tau$ with $1/\dot{\gamma}$ at $\delta U \sim U$. Empirically, we find St $\sim \mathcal{O}(10^{-2})$ is often sufficient for the particle inertia to be negligible.
  
\item A vanishing stiffness-scaled shear rate, $\hat{\dot{\gamma}}=\dot{\gamma}a_1/\sqrt{k_n/(\rho a_1)} \ll 1$.

  $\hat{\dot{\gamma}}$ describes the particle ``hardness''. It follows from the scaling analysis that two particles of same radius $a$, at overlap distance $\delta$, have a contact area $A \approx \pi a \delta$. Given the collision force $F \approx k_n \delta$, the characteristic velocity is then $U \sim \sqrt{P/\rho} \sim \sqrt{k_n/(\rho \pi a)}$. This provides a collision time scale $\tau \sim a/\sqrt{k_n/(\rho \pi a)}$ that can be compared with $1/\dot{\gamma}$. Empirically, we find $\hat{\dot{\gamma}} \sim \mathcal{O}(10^{-4})$ is sufficient for the particles to be considered as hardspheres.
  
\item A vanishing non-dimensional relaxation time, $\hat{\tau}=\gamma_n\dot{\gamma}/k_t$ or $\mu a_1 \dot{\gamma}/k_t \ll 1$.

  $\hat{\tau}$ describes the non-dimensional relaxation time associated with a contact. Both normal and tangential contacts have relaxation times; here, we require the latter to be much less than 1, since $k_t < k_n$ typically. The second definition for $\hat{\tau}$ above uses the normal lubrication force as the dashpot in case $\gamma_n=0$, cf.\ Eqs.\ (\ref{lub F}, \ref{eq:lub79}, \ref{eq:lub97}).
  
\end{enumerate}

Apart from the above conditions, the behaviour of the simulated suspension is determined by the volume fraction $\phi$ and any force-rescaled shear rate $\hat{\dot{\gamma}}_r=6\pi\mu a_1^2/F$ if an additional non-hydrodynamic force scale exists in the system, \eg friction or electrostatic repulsion. Physically, $\hat{\dot{\gamma}}_r$ is usually introduced to invoke a rate-dependent rheology.

\subsection{Stress tensor and bulk rheology}
\label{sec:stress}

The bulk stress tensor is calculated as
\begin{equation} 
  \begin{aligned}
    \mathbb{\Sigma} = 2\mu \mathbb{E}^\infty + \frac{1}{V} 
    \bigg(\sum_i^N\mathbb{S}_i^S +\sum_{i,j}^{N_l}\mathbb{S}_{ij}^L +\sum_{i,j}^{N_c}\mathbb{S}_{ij}^C
    +\sum_{i,j}^{N_r}\mathbb{S}_{ij}^R+\sum_{i,j}^{N_a}\mathbb{S}_{ij}^A \bigg), 
  \end{aligned}
  \label{b stress}
\end{equation} 
where $V=L_xL_yL_z$ is the volume of the simulation box, and $\mathbb{S}$'s denote the stresslets due to various interactions. In  the conventional coordinate system where $x$, $y$, $z$ denote the streamwise, velocity-gradient, and vorticity directions, respectively, the individual terms in Eq.\ \eqref{b stress} corresponding to a simple shear flow are given by 
\begin{subequations}
  \begin{equation} 
    \begin{aligned}
      \mathbb{E}^\infty = 
      \begin{bmatrix} 
        0&\nicefrac{\dot{\gamma}}{2}&0 \\ \nicefrac{\dot{\gamma}}{2}&0&0  \\ 0&0&0 
      \end{bmatrix}, 
    \end{aligned}
  \end{equation}
  \begin{equation} 
    \begin{aligned}
      \mathbb{S}_i^S = (20\pi \mu a_{i}^3/3) \mathbb{E}^\infty,
    \end{aligned}
    \label{stresslet A}
  \end{equation}
  \begin{equation} 
    \begin{aligned}
      \mathbb{S}_{mn}^L = (F_m^L r_n + F_n^Lr_m)/2,
    \end{aligned}
    \label{stresslet L}
  \end{equation}
  \begin{equation} 
    \begin{aligned}
      \mathbb{S}_{mn}^C = F_m^C r_n,
    \end{aligned}
    \label{stresslet C}
  \end{equation}
\end{subequations}
where ${\bm r}$ is the separation vector pointing from particle $i$ to particle $j$, and Eq.\ \eqref{stresslet C} applies to other interparticle forces as well. Note that, the calculation of the lubrication stresslet, Eq.\ \eqref{stresslet L}, involves a simplification: the $\mathcal{O}(\delta)$ and isotropic terms are neglected; see \cite{ranga} for the derivation. Note also that, the subscript in Eq.\ \eqref{stresslet A} refers to particle $i$, while the subscripts in Eqs.\ (\ref{stresslet L}--\ref{stresslet C}) denotes the tensor notation.

Once the stress tensor is obtained, the shear stress $\sigma$, normal stress differences $N_1$ and $N_2$, and particle pressure $\Pi$ can be readily calculated from the following definitions,
\begin{subequations}
  \begin{equation} 
    \begin{aligned}
      \sigma=\mathbb{\Sigma}_{12},
    \end{aligned}
  \end{equation}
  \begin{equation} 
    \begin{aligned}
      N_1=\mathbb{\Sigma}_{11}-\mathbb{\Sigma}_{22}, \quad
      N_2=\mathbb{\Sigma}_{22}-\mathbb{\Sigma}_{33},
    \end{aligned}
  \end{equation}
  \begin{equation} 
    \begin{aligned}
      \Pi=-\textrm{Tr}(\mathbb{\Sigma})/3.
    \end{aligned}
  \end{equation}
\end{subequations}
The relative viscosity and non-dimensional particle pressure are defined as $\eta_r=\sigma/(\mu\dot{\gamma})$ and $\eta_n=\Pi/(\mu\dot{\gamma})$, respectively.

\subsection{Numerical integration}
\label{sec:numerical}

The governing equations presented in Sec.\ \ref{sec:eqs} can be integrated in time explicitly using the modified velocity-Verlet algorithm \citep{Groot_Warren_1997},
\begin{subequations}
  \begin{equation} 
    \begin{aligned}
      {\bm x}_i^{(n+1)} = {\bm x}_i^{(n)}+\Delta t {\bm v}_i^{(n)} + \frac{\Delta t^2}{2} {\bm \alpha}_i^{(n)},
    \end{aligned} \label{verlect 1}
  \end{equation}
  \begin{equation} 
    \begin{aligned}
      {\bm v}_i^{(n+1/2)} = {\bm v}_i^{(n)}+ \frac{\Delta t}{2} {\bm \alpha}_i^{(n)},
    \end{aligned} \label{verlect 2}
  \end{equation}
  \begin{equation} 
    \begin{aligned}
      {\bm \alpha}_i^{(n+1)} = \mathcal{F} \bigg\{{\bm x}_i^{(n+1)}, {\bm v}_i^{(n+1/2)} \bigg\},
    \end{aligned} \label{verlect 3}
  \end{equation}
  \begin{equation} 
    \begin{aligned}
      {\bm v}_i^{(n+1)} = {\bm v}_i^{(n)}+ \frac{\Delta t}{2} \bigg({\bm \alpha}_i^{(n)} + {\bm \alpha}_i^{(n+1)} \bigg),
    \end{aligned} \label{verlect 4}
  \end{equation}
\end{subequations}
where ${\bm x}_i^{(n)}=(x,y,z)_i^{(n)}$, ${\bm v}_i^{(n)}=(u,v,w)_i^{(n)}$, and ${\bm \alpha}_i^{(n)}$ denote the position, velocity, and acceleration vectors of particle $i$, respectively, at time $t=n\Delta t$, and $\mathcal{F}$ denotes the force functional as in Eq.\ \eqref{force-decomp}. For spherical particles, the same update procedure applies to particle orientations, angular velocities, and angular accelerations (see Sec.\ \ref{sec:nonsph} for non-spherical particles).

To comply with simple shear flows at fixed volume, the Lees-Edwards boundary condition is imposed on particle positions and their $u$ velocity components to remove the wall effects and reduce the size of the computational box \citep{Lees_Edwards_1972}. Physically, it results in a homogeneous suspension with a net momentum flux in the $y$ direction if the system is far from equilibrium, and it reads
\begin{subequations}
  \begin{equation} 
    \begin{aligned}
      x=
      \begin{cases}
        (x+L_x-x_{sh}) \mod L_x & \textrm{if} \quad  y>L_y, \\
        (x+L_x+x_{sh}) \mod L_x & \textrm{if} \quad  y<0, \\
        (x+L_x) \mod L_x & \textrm{otherwise}, \\
      \end{cases}
    \end{aligned}
  \end{equation}
  \begin{equation} 
    \begin{aligned}
      y=(y+L_y) \mod L_y,
    \end{aligned}
  \end{equation}
  \begin{equation} 
    \begin{aligned}
      z=(z+L_z) \mod L_z,
    \end{aligned}
  \end{equation}
  \begin{equation} 
    \begin{aligned}
      u=
      \begin{cases}
        u-u_{sh}  & \textrm{if} \quad  y>L_y, \\
        u+u_{sh}  & \textrm{if} \quad  y<0, 
      \end{cases}
    \end{aligned}
  \end{equation}
\end{subequations}
where
\begin{equation} 
  \begin{aligned}
    x_{sh} = \dot{\gamma}L_yt \mod L_x, \quad u_{sh} = \dot{\gamma}L_y,
  \end{aligned}
\end{equation}
define the position and velocity shifts at time $t$.

The above explicit integration scheme is second-order accurate in time and requires the time step $\Delta t$ to be smaller than the smallest physical time scale of the process, cf.\ Sec.\ \ref{model param}. Empirical experience suggests $\Delta t \dot{\gamma} \approx 10^{-6}$. Alternatively, the equations of motion can be solved by matrix inversion upon setting the left-hand-side of Eq.\ \eqref{newton-euler} to zero, thus obtaining particle velocities in the quasi-static limit \citep{Mari_Seto_2014JoR}. The latter approach is equivalent to the one adopted here provided St $\ll 1$, see \eg \cite{Ness_Mari_Cates_2018}.

Finally, we note that the force calculation, Eq.\ \eqref{verlect 3}, is usually the most time-consuming step in a Verlet integration. One straightforward technique to speed up the computation is to construct a near-neighbour list (NNL) for each particle and only calculate the force between particle pairs therein. This way, the operation count for a complete update reduces from being quadratic to being linear with the number of particles.\footnote{The overhead due to the construction of a NNL normally does not exceed the cost reduction for dense suspensions.} Practically, the algorithm requires $\mathcal{O}(cN)$ operations, where $c$ is a kernel constant mainly depending on the cutoff distance of particle interactions.\footnote{The lubrication cutoff is usually chosen as 0.05min$(a_1,a_2)$ for dense suspensions, see \cite{Cheal_Ness_2018}.}

\section{Validations}
\label{valid}

In this section, we validate our numerical implementation of the HLGD model with two benchmark rheologies, \ie jamming transition and shear thickening.

\subsection{Jamming transition}

We simulate 200 initially randomly distributed spherical particles in a cubic box. A monodisperse suspension is consider for the lowest volume fraction $\phi=40$\%, while bidisperse suspensions with radius ratio $a_2/a_1=1.4$ in equal volumes are used for higher volume fractions to prevent ordering. The random seeds are generated by the protocol of Mari and Seto.\footnote{See \url{https://github.com/rmari/LF_DEM} for details.} The particles interact via Stokes drag, lubrication, and collision forces without any friction or physico-chemical interactions. The parameters used for this case are summarized in Table \ref{tab:param}.\footnote{Numerical convergence is checked by halving $\dot{\gamma}$ or doubling $k_n$, where roughly the same results are obtained.}

\begin{table}[h]
  \centering
  \caption{Summary of parameters.}
  \renewcommand{\arraystretch}{1.2}
  \begin{tabular}{ccccccc}
    \hline
    $N$ & St & $\hat{\dot{\gamma}}$ & $\hat{\tau}$ & $\mu_c$ & $t\dot{\gamma}$\\
    \hline
    200 & $10^{-2}\sim 10^{-3}$ & $7\times 10^{-6}\sim 10^{-4}$ & $5\times 10^{-8}\sim 10^{-6}$  & 0 & 10 \\
    \hline
  \end{tabular}
  \label{tab:param}
\end{table}

\begin{figure}
  \centering
  \includegraphics[width=0.5\columnwidth]{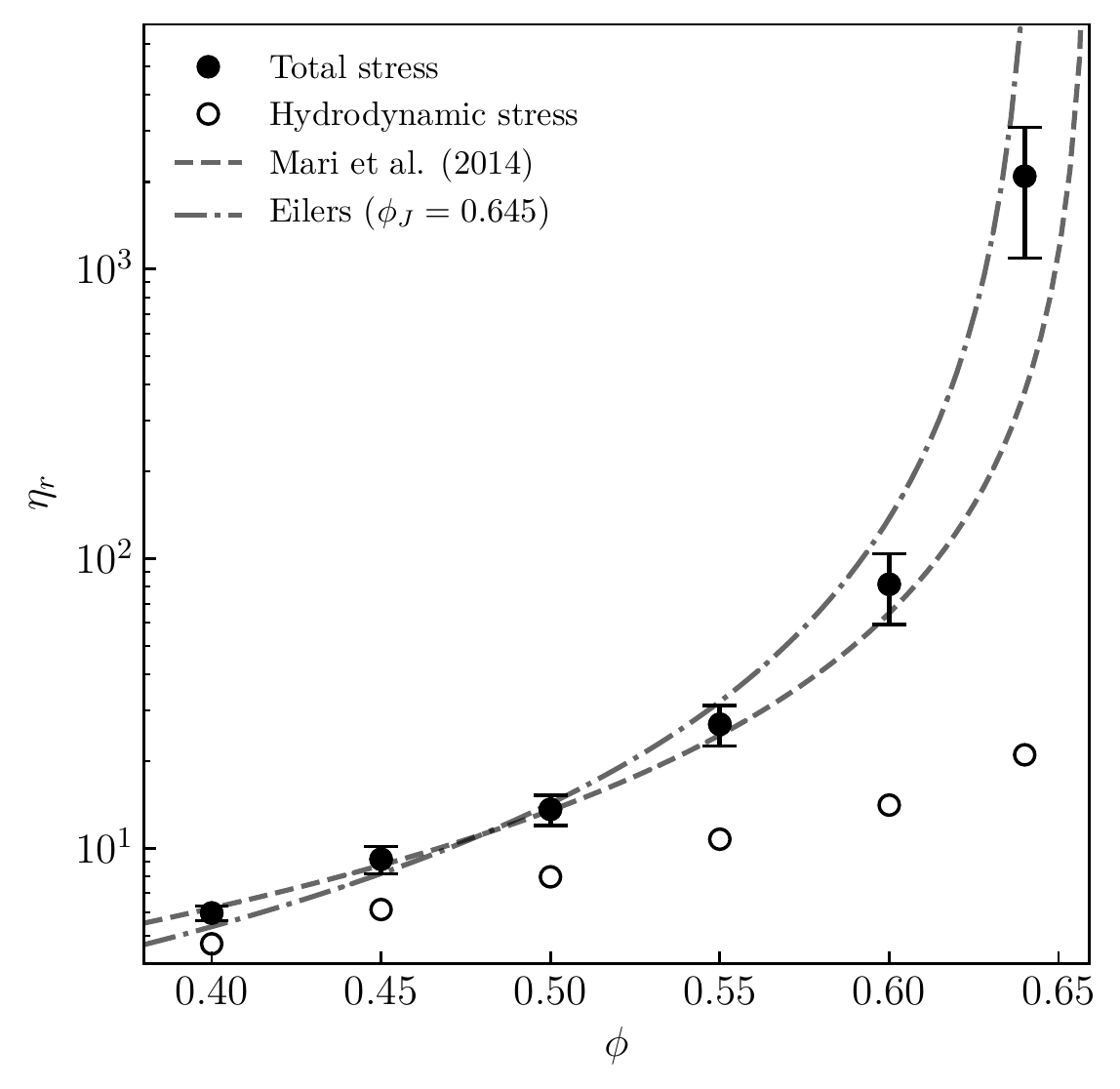}
  \caption{Relative viscosity versus volume fraction for suspensions of frictionless spheres.}
  \label{fig:visc}
\end{figure}

Figure \ref{fig:visc} shows the relative viscosity for six volume fractions ranging from 40\% to 64\%. The data pertain temporal averages and standard deviations calculated in the standard way over a strain of 10. Clearly, a jamming transition indicated by a diverging $\eta_r$ as $\phi$ increases is observed in favorable comparison with both previous numerical simulations and an empirical correlation. Specifically, the power-law fitting of \cite{Mari_Seto_2014JoR} satisfies $\eta_r=1.4(1-\phi/\phi_J)^{-1.6}$, with the jamming volume fraction $\phi_J=0.66$; while the Eilers' correlation has the form $\eta_r=[1+(5\phi/4)/(1-\phi/\phi_J)]^2$, where we take $\phi_J=0.645$.  Our data fall well within the two limits, suggesting a jamming volume fraction at approximately 65\% for frictionless spheres. Note that, the exact value of $\phi_J$ is sensitive to the particle overlap (numerically) or surface roughness (experimentally), as the contact stress dominates over the hydrodynamic one at high volume fractions. For spheres, the latter becomes less than 10\% of the total stress for $\phi \gtrapprox 0.6$, see Figure \ref{fig:visc}.

\begin{figure}
  \centering
  \includegraphics[width=0.5\columnwidth]{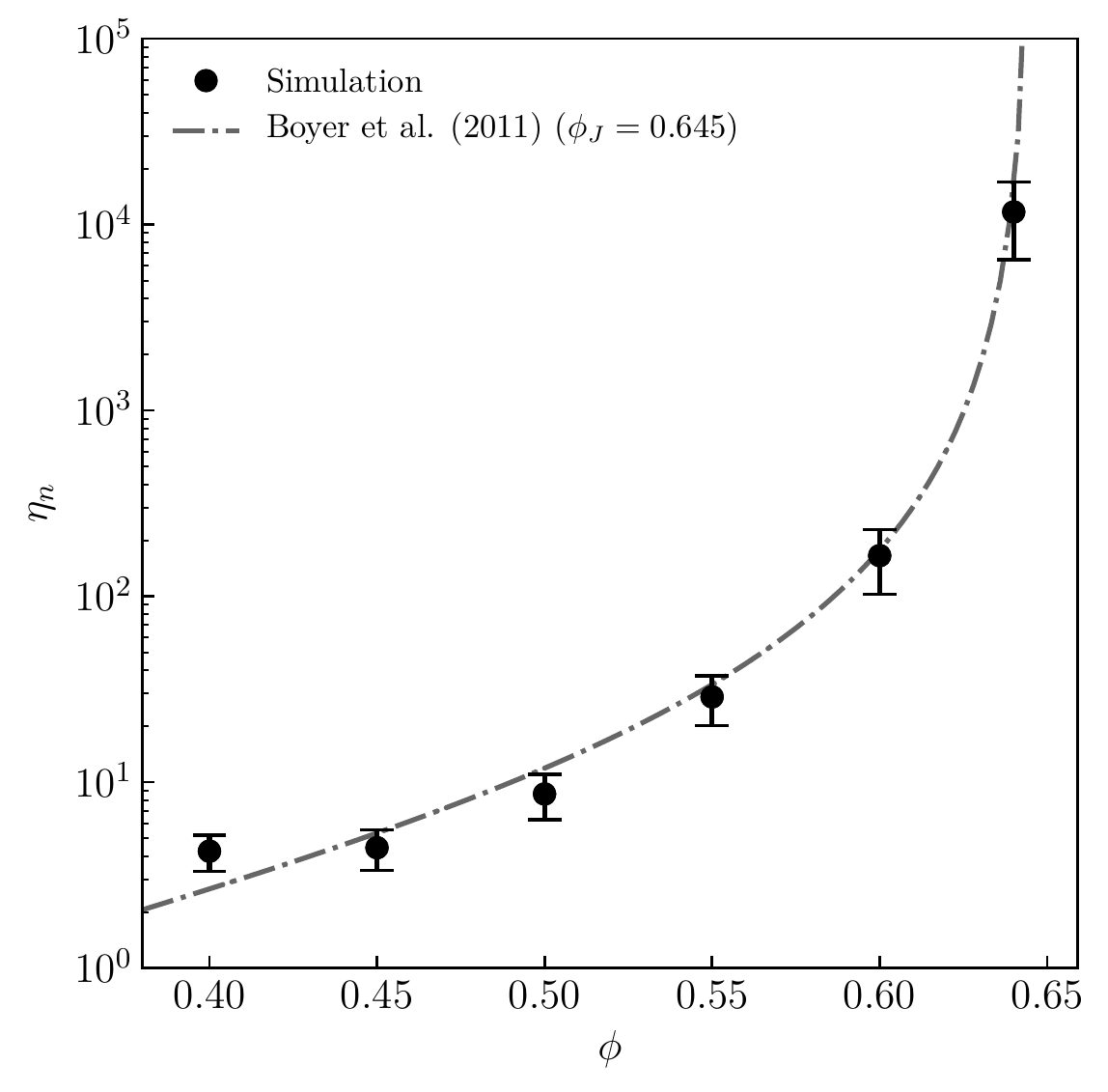}
  \caption{Particle pressure (normalized) versus volume fraction for suspensions of frictionless spheres.}
  \label{fig:pres}
\end{figure}

We also examine the non-dimensional particle pressure of the same suspension and compare our results with the theoretical model of \cite{Boyer_Guaz_Poul_2011}. As illustrated in Figure \ref{fig:pres}, an even steeper divergence of $\eta_n$ with $\phi$ is observed, suggestive of a strong tendency for the system to dilate. The constitutive law derived by \cite{Boyer_Guaz_Poul_2011}, supported by their own experiment, is given as $\eta_n=[\phi/(\phi_J-\phi)]^2$. Taking $\phi_J=0.645$, an overall excellent agreement is observed except for the lowest volume fraction $\phi=40$\%, where our data slightly over-predicts the particle pressure. In general, the comparisons above verify our implementation of the HLGD model in the case of jamming transition of frictionless spheres.

\subsection{Shear thickening}

\begin{figure}
  \centering
  \includegraphics[width=0.5\columnwidth]{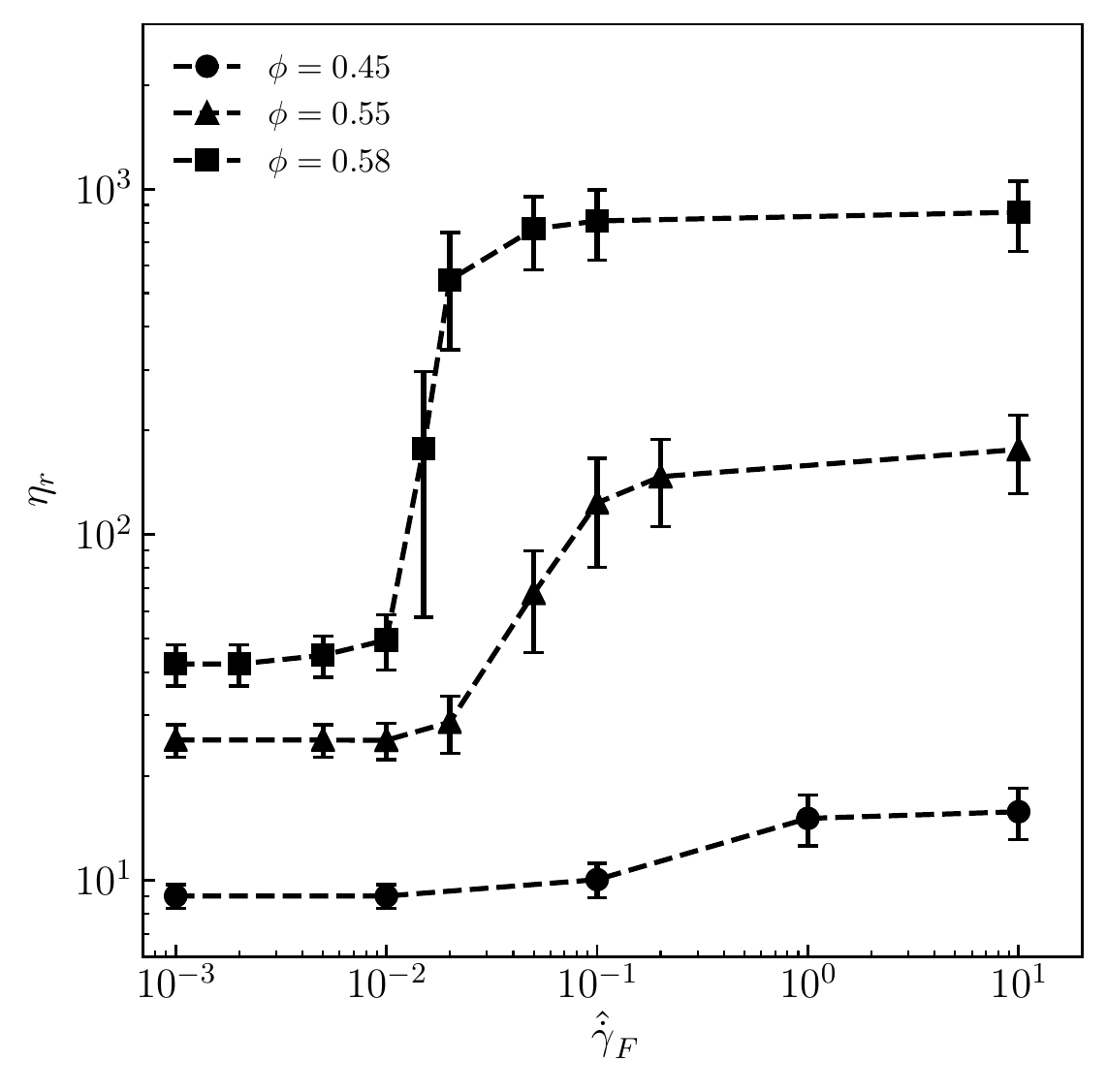}
  \caption{Continuous and discontinuous thickening of suspensions of frictional spheres. $N=500$, $\mu_c=0.5$.}
  \label{fig:thick}
\end{figure}

As contact becomes more important at higher particle concentrations, we simulate three frictional suspensions displaying continuous and discontinuous shear thickening rheologies. Specifically, we implement the critical-load friction model introduced in \cite{Mari_Seto_2014JoR} that activates the friction force only when the normal collision force exceeds a critical value. Figure \ref{fig:thick} reproduces the results of \cite{Mari_Seto_2014JoR}. Here, the relative viscosity is plotted against a non-dimensional shear rate defined by the ratio of the Stokes drag and the threshold friction, cf.\ Sec.\ \ref{model param}. At 45\% volume fraction, the suspension viscosity increases gradually with the shear due to increased particle contact. The slope steepens for $\phi=0.55$; while at $\phi=0.58$, we observe an abrupt increase of the mean relative viscosity, accompanied by large fluctuations, indicating a switch from CST to DST. Accurate prediction of the onset of DST remains a theoretical challenge \citep{Morris_annurev2020}. Our results demonstrate the present implementation of the HLGD model can be used to probe such rheologies in details.


\section{Extension to non-spherical particles}
\label{sec:nonsph}

Finally, we turn our attention to suspensions of non-spherical particles. While spheres represent the simplest geometry convenient for theoretical and numerical studies, suspensions in reality are almost certainly composed of non-spherical particles. For very dilute suspensions, this may not be an important issue as the macroscopic behaviours are usually not very sensitive to the exact particle shape. However, for dense suspension, a completely different rheology or phase transition may be expected as the lubrication intensifies and particle contact increases, see \eg \cite{Damasceno453, Royer_etal_2015, trulsson_2018}. To simulate suspensions in the latter case, various levels of simplifications may be made depending on the specific particle shape and volume fraction. Here, we present a minimal model that captures the essential effects, mostly suitable for spherical aggregates or polyhedra at high concentrations, extended from the HLGD model.

Assuming collisions and frictions are the most significant factor to the suspension rheology, we can approximate the hydrodynamic interactions using the same functional forms as presented in Sec.\ \ref{sec:eqs}. That is, each particle experiences a Stokes drag based on some hydraulic radius, $\hat{a}$, and each neighbouring pair interacts via the lubrication force as if they were spheres. The exact value of $\hat{a}$ depends on the shape of the particle, \eg $\hat{a}=L/2$ for a cube of length $L$; in general, this is an approximation that shall not greatly affect the bulk behaviour. In a similar spirit, the radii used in the lubrication calculation can be taken as $\hat{a}$ or the inverse of the local curvature. We note that more rigorous treatments of the hydrodynamic interactions have been proposed, though it is rather complicated to implement and is limited to spheroids, see \eg \cite{claeys_brady_1993}. Our simplified approach is similar to the mean-field description developed for wet foams, which has also been employed in studies of frictionless particles \citep{Durian_1997PRE, Marschall_etal_2019}.

For non-spherical particles, an additional complication arises due to the generally nontrivial rigid-body dynamics, where particle orientations matter and the moment-of-inertia, ${\bm I}_i$, is a tensor. This motivates us to use a quaternion-based, predictor-corrector direct multiplication (PCDM) scheme, as follows
\begin{subequations}
  \begin{equation} 
    \begin{aligned}
      {\bm x}_i^{(n+1)} = {\bm x}_i^{(n)}+\Delta t {\bm v}_i^{(n+1/2)},
    \end{aligned} \label{qua 1}
  \end{equation}
  \begin{equation} 
    \begin{aligned}
      {\bm v'}_i^{(n+1)} = {\bm v}_i^{(n+1/2)}+ \frac{\Delta t}{2} {\bm \alpha}_i^{(n)},
    \end{aligned} \label{qua 2a}
  \end{equation}
  \begin{equation} 
    \begin{aligned}
      {\bm \omega}_i^{(n+3/4),b} = {\bm \omega}_i^{(n+1/2),b} + \frac{\Delta t}{4} {\bm \beta}_i^{(n),b},
    \end{aligned} \label{qua 2b1}
  \end{equation}
  \begin{equation} 
    \begin{aligned}
      {\bm \omega}_i^{(n+3/4)} = q^{(n+1/2)} {\bm \omega}_i^{(n+3/4),b} \bigg( q^{(n+1/2)} \bigg)^{-1},
    \end{aligned} \label{qua 2b2}
  \end{equation}
  \begin{equation} 
    \begin{aligned}
      {q'}^{(n+1)} =\bigg[ \cos \bigg( \frac{ ||{\bm \omega}_i^{(n+3/4)}|| \Delta t}{4} \bigg),
        \sin \bigg( \frac{ ||{\bm \omega}_i^{(n+3/4)}|| \Delta t}{4} \bigg)
        \frac{{\bm \omega}_i^{(n+3/4)}}{||{\bm \omega}_i^{(n+3/4)}||}  \bigg] q^{(n+1/2)},
    \end{aligned} \label{qua 2b3}
  \end{equation}
  \begin{equation} 
    \begin{aligned}
      {\bm \omega '}_i^{(n+1),b} = {\bm \omega}_i^{(n+1/2),b} + \frac{\Delta t}{2} {\bm \beta}_i^{(n),b},
    \end{aligned} \label{qua 2c1}
  \end{equation}
  \begin{equation} 
    \begin{aligned}
      {\bm \omega '}_i^{(n+1)} = {q'}^{(n+1)} {\bm \omega '}_i^{(n+1),b} \bigg( {q'}^{(n+1)} \bigg)^{-1},
    \end{aligned} \label{qua 2c2}
  \end{equation}
  \begin{equation} 
    \begin{aligned}
      {\bm \alpha}_i^{(n+1)} = \mathcal{F} \bigg\{{\bm x}_i^{(n+1)}, {\bm v'}_i^{(n+1)}, {q'}^{(n+1)}, {\bm \omega '}_i^{(n+1)} \bigg\},
    \end{aligned} \label{qua 3a}
  \end{equation}
  \begin{equation} 
    \begin{aligned}
      {\bm T}_i^{(n+1)} = \mathcal{T} \bigg\{{\bm x}_i^{(n+1)}, {\bm v'}_i^{(n+1)}, {q'}^{(n+1)}, {\bm \omega '}_i^{(n+1)} \bigg\},
    \end{aligned} \label{qua 3b1}
  \end{equation}
  \begin{equation} 
    \begin{aligned}
      {\bm T}_i^{(n+1),b} = \bigg( {q'}^{(n+1)} \bigg)^{-1} {\bm T}_i^{(n+1)}  {q'}^{(n+1)},
    \end{aligned} \label{qua 3b2}
  \end{equation}
  \begin{equation} 
    \begin{aligned}
      {\bm \beta}_i^{(n+1),b} = \bigg( {\bm I}_i^{b} \bigg)^{-1}
      \bigg( {\bm T}_i^{(n+1),b} - {\bm \omega '}_i^{(n+1),b} \times {\bm I}_i^{b}{\bm \omega '}_i^{(n+1),b} \bigg)  ,
    \end{aligned} \label{qua 3b3}
  \end{equation}
  \begin{equation} 
    \begin{aligned}
      {\bm \omega}_i^{(n+3/2),b} = {\bm \omega}_i^{(n+1/2),b} + \Delta t {\bm \beta}_i^{(n+1),b},
    \end{aligned} \label{qua 4a}
  \end{equation}
  \begin{equation} 
    \begin{aligned}
      {q}^{(n+3/2)} =\bigg[ \cos \bigg( \frac{ ||{\bm \omega '}_i^{(n+1)}|| \Delta t}{2} \bigg),
        \sin \bigg( \frac{ ||{\bm \omega '}_i^{(n+1)}|| \Delta t}{2} \bigg)
        \frac{{\bm \omega '}_i^{(n+1)}}{||{\bm \omega '}_i^{(n+1)}||}  \bigg] q^{(n+1/2)},
    \end{aligned} \label{qua 4b}
  \end{equation}
  \begin{equation} 
    \begin{aligned}
      {\bm \omega}_i^{(n+3/2)} = q^{(n+3/2)} {\bm \omega}_i^{(n+3/2),b} \bigg( q^{(n+3/2)} \bigg)^{-1},
    \end{aligned} \label{qua 4c}
  \end{equation}
  \begin{equation} 
    \begin{aligned}
      {\bm v}_i^{(n+3/2)} = {\bm v}_i^{(n+1/2)}+ \Delta t {\bm \alpha}_i^{(n+1)},
    \end{aligned} \label{qua 4d}
  \end{equation}
\end{subequations}
where $q=q_0+{\bm q}$ is the quaternion, ${\bm \beta}$ is the angular acceleration, and superscripts prime and $b$ denote prediction and body-frame values, respectively. Clearly, the integration scheme becomes much more cumbersome when the particles are non-spherical.

For collision detection, we employ the classical GJK algorithm that efficiently computes the Euclidean distance between a pair of convex sets \citep{GJK1988}. The resulting bulk rheology can be calculated in the same way as in Sec.\ \ref{sec:stress}, \ie summing the first moment of various force fields. Validation of the present algorithm will be presented in upcoming publications.

\section*{Acknowledgments}

Detailed discussions with Romain Mari, Christopher Ness, and Rangarajan Radhakrishnan on the numerics are gratefully acknowledged. This work is supported by the Microflusa project that receives funding from the European Union Horizon 2020 research and innovation programme under Grant Agreement No.\ 664823.


\section*{Appendix}

\renewcommand\thesubsection{\Alph{subsection}}  
\counterwithin*{equation}{subsection}  
\renewcommand{\theequation}{\thesubsection.\arabic{equation}} 

\subsection{Lubrication resistances} \label{app: scalar resist}

The scalar resistances introduced in Eqs.\ (\ref{lub F}--\ref{lub T 2}) follow those in \cite{kim_karrila,Cheal_Ness_2018}, and are given as
\begin{subequations}
  \begin{equation}  \label{eq:lub79}
    \begin{aligned}
      X_{ii}^A = a_i (R_{xiia1}/\delta + R_{xiia2} \log(1/\delta)),
    \end{aligned}
  \end{equation}
  \begin{equation} 
    \begin{aligned}
      Y_{ii}^A = a_i  R_{yiia} \log(1/\delta),
    \end{aligned}
  \end{equation}
  \begin{equation} 
    \begin{aligned}
      Y_{ii}^B = a_i^2  R_{yiib} \log(1/\delta),
    \end{aligned}
  \end{equation}
  \begin{equation} 
    \begin{aligned}
      Y_{ji}^B = a_j^2  R_{yjib} \log(1/\delta),
    \end{aligned}
  \end{equation}
  \begin{equation} 
    \begin{aligned}
      Y_{ii}^C = a_i^3  R_{yiic} \log(1/\delta),
    \end{aligned}
  \end{equation}
  \begin{equation} 
    \begin{aligned}
      Y_{ij}^C = Y_{ji}^C = a_i^3  R_{yijc} \log(1/\delta),
    \end{aligned}
  \end{equation}
  \begin{equation} 
    \begin{aligned}
      Y_{jj}^C = a_j^3  R_{yjjc} \log(1/\delta),
    \end{aligned}
  \end{equation}
\end{subequations}
where $\delta=2h_{ij}/(a_i+a_j)$ is the non-dimensional surface gap between particles $i$ and $j$, and the coefficients $R$'s are functions of the size ratio $\lambda=a_j/a_i$ and $\mu$. Specifically, they are calculated as

\begin{subequations}
  \begin{equation}  \label{eq:lub97}
    \begin{aligned}
      R_{xiia1} = (6\pi \mu)\frac{2\lambda^2}{(1+\lambda)^3}, \quad
      R_{xiia2} = (6\pi \mu)\frac{\lambda (1+7\lambda+\lambda^2)}{5(1+\lambda)^3},
    \end{aligned}
  \end{equation}
  \begin{equation} 
    \begin{aligned}
      R_{yiia} = (6\pi \mu)\frac{4\lambda (2+\lambda+2\lambda^2)}{15(1+\lambda)^3},
    \end{aligned}
  \end{equation}
  \begin{equation} 
    \begin{aligned}
      R_{yiib} = (-4\pi \mu)\frac{\lambda (4+\lambda)}{5(1+\lambda)^2},
    \end{aligned}
  \end{equation}
  \begin{equation} 
    \begin{aligned}
      R_{yjib} = (-4\pi \mu)\frac{\lambda^{-1} (4+\lambda^{-1})}{5(1+\lambda^{-1})^2},
    \end{aligned}
  \end{equation}
  \begin{equation} 
    \begin{aligned}
      R_{yiic} = (8\pi \mu)\frac{2\lambda}{5(1+\lambda)},
    \end{aligned}
  \end{equation}
  \begin{equation} 
    \begin{aligned}
      R_{yijc} = (8\pi \mu)\frac{\lambda^2}{10(1+\lambda)}.
    \end{aligned}
  \end{equation}
  \begin{equation} 
    \begin{aligned}
      R_{yjjc} = (8\pi \mu)\frac{2\lambda^{-1}}{5(1+\lambda^{-1})}.
    \end{aligned}
  \end{equation}
\end{subequations}

\subsection{Numerical treatment of the lubrication at small and large distances}  \label{app: lub range}

We note that the lubrication force and torque are singular at contact, with $X \sim 1/\delta$ and $Y \sim \log (1/\delta)$, see Appendix \ref{app: scalar resist}. The singularity derives from assuming perfectly smooth particles in the mathematical sense. In practice, particle contact is inevitable due to surface roughness; thus, we allow small overlap to occur invoking both lubrication and contact forces. Numerically, the lubrication force is saturated below $h_{inner}=0.001a_1$ and truncated above $h_{outer}=(0.05\sim 0.2)a_1$, where $a_1$ denotes the smallest particle radius. The outer range is introduced to reduce the computational cost and is adjusted by examining the radial distribution function at each volume fraction.\footnote{The lubrication formulation has been simplified comparing to \cite{jeffrey_onishi_1984,jeffrey1992} by neglecting terms of $\mathcal{O}(1)$ or higher order. Therefore, the lubrication outer cutoff cannot be arbitrarily large. See \cite{ranga} for the detailed algebra.}

\subsection{Tangential stretch for contacting particles} \label{app: tang stretch}

Following Appendix \ref{app: lub range}, another consequence of the particle roughness is the initiation of frictional contact. Here, we adopt the standard stick/slide model for the calculation of the friction force, as given in Eqs.\ (\ref{col F}--\ref{col F 2}) \citep{Cundall_Strack1979,Luding2008}. Specifically, the tangential stretch vector is calculated as
\begin{equation} 
  \begin{aligned}
    {\bm \xi}_{ij}(t)= 
    \begin{cases}
      \int_{t_0}^t -P_t
      [({\bm u}_i-{\bm u}_j)+(a_i{\bm \omega}_i+a_j{\bm \omega}_j) \times {\bm n}_{ij}] dt', 
      & \text{if } \abs{{\bm \xi}_{ij}} < \abs{{\bm \xi}_{max}}, \\
      {\bm \xi}_{max} , & \text{otherwise},
    \end{cases}
  \end{aligned}
  \label{tang stretch}
\end{equation}
where ${\bm \xi}_{max}$ is maximal tangential stretch allowed by the Coulomb's law of friction (Eq.\ \ref{col F 2}), and $t_0$ is the moment a frictional contact is established. We choose $k_t=(2/7)k_n$ as given in \cite{Cheal_Ness_2018}.

\bibliographystyle{jfm}
\bibliography{main}

\begin{thebibliography}{32}
\expandafter\ifx\csname natexlab\endcsname\relax\def\natexlab#1{#1}\fi

\bibitem[Alder \& Wainwright(1959)]{Alder_Wainwright_1959}
{\sc Alder, B.~J. \& Wainwright, T.~E.} 1959 Studies in molecular dynamics.
  {I}. {G}eneral method. {\em The Journal of Chemical Physics\/} {\bf 31}~(2),
  459--466.

\bibitem[Batchelor(1967)]{Batchelor}
{\sc Batchelor, G.K.} 1967 {\em An introduction to fluid dynamics\/}. Cambridge
  University Press.

\bibitem[Batchelor \& Green(1972)]{batchelor_green_1972b}
{\sc Batchelor, G.~K. \& Green, J.~T.} 1972 The determination of the bulk
  stress in a suspension of spherical particles to order c2. {\em Journal of
  Fluid Mechanics\/} {\bf 56}~(3), 401–427.

\bibitem[Boyer {\em et~al.\/}(2011)Boyer, Guazzelli \&
  Pouliquen]{Boyer_Guaz_Poul_2011}
{\sc Boyer, Fran\ifmmode \mbox{\c{c}}\else~\c{c}\fi{}ois, Guazzelli,
  \'Elisabeth \& Pouliquen, Olivier} 2011 Unifying suspension and granular
  rheology. {\em Phys. Rev. Lett.\/} {\bf 107}, 188301.

\bibitem[Brady \& Bossis(1988)]{Brady_Bossis1988}
{\sc Brady, J~F \& Bossis, G} 1988 Stokesian dynamics. {\em Annual Review of
  Fluid Mechanics\/} {\bf 20}~(1), 111--157.

\bibitem[Cheal \& Ness(2018)]{Cheal_Ness_2018}
{\sc Cheal, Oliver \& Ness, Christopher} 2018 Rheology of dense granular
  suspensions under extensional flow. {\em Journal of Rheology\/} {\bf 62}~(2),
  501--512.

\bibitem[Claeys \& Brady(1993)]{claeys_brady_1993}
{\sc Claeys, Ivan~L. \& Brady, John~F.} 1993 Suspensions of prolate spheroids
  in stokes flow. part 1. dynamics of a finite number of particles in an
  unbounded fluid. {\em Journal of Fluid Mechanics\/} {\bf 251}, 411–442.

\bibitem[Cundall \& Strack(1979)]{Cundall_Strack1979}
{\sc Cundall, P.~A. \& Strack, O. D.~L.} 1979 {A discrete numerical model for
  granular assemblies.} {\em Geotechnique\/} {\bf 29}~(1), 47--65.

\bibitem[Damasceno {\em et~al.\/}(2012)Damasceno, Engel \&
  Glotzer]{Damasceno453}
{\sc Damasceno, Pablo~F., Engel, Michael \& Glotzer, Sharon~C.} 2012 Predictive
  self-assembly of polyhedra into complex structures. {\em Science\/} {\bf
  337}~(6093), 453--457.

\bibitem[Denn \& Morris(2014)]{Morton_Morris_2014}
{\sc Denn, Morton~M. \& Morris, Jeffrey~F.} 2014 Rheology of non-brownian
  suspensions. {\em Annual Review of Chemical and Biomolecular Engineering\/}
  {\bf 5}~(1), 203--228, pMID: 24655134.

\bibitem[Durian(1997)]{Durian_1997PRE}
{\sc Durian, D.~J.} 1997 Bubble-scale model of foam mechanics:mmelting,
  nonlinear behavior, and avalanches. {\em Phys. Rev. E\/} {\bf 55},
  1739--1751.

\bibitem[Einstein(1906)]{Einstein_1906}
{\sc Einstein, A.} 1906 Eine neue bestimmung der moleküldimensionen. {\em
  Annalen der Physik\/} {\bf 324}~(2), 289--306.

\bibitem[{Gilbert} {\em et~al.\/}(1988){Gilbert}, {Johnson} \&
  {Keerthi}]{GJK1988}
{\sc {Gilbert}, E.~G., {Johnson}, D.~W. \& {Keerthi}, S.~S.} 1988 A fast
  procedure for computing the distance between complex objects in
  three-dimensional space. {\em IEEE Journal on Robotics and Automation\/} {\bf
  4}~(2), 193--203.

\bibitem[Groot \& Warren(1997)]{Groot_Warren_1997}
{\sc Groot, Robert~D. \& Warren, Patrick~B.} 1997 Dissipative particle
  dynamics: Bridging the gap between atomistic and mesoscopic simulation. {\em
  The Journal of Chemical Physics\/} {\bf 107}~(11), 4423--4435.

\bibitem[Guazzelli \& Pouliquen(2018)]{guazzelli_pouliquen_2018}
{\sc Guazzelli, Élisabeth \& Pouliquen, Olivier} 2018 Rheology of dense
  granular suspensions. {\em Journal of Fluid Mechanics\/} {\bf 852}, P1.

\bibitem[Hoogerbrugge \& Koelman(1992)]{Hoogerbrugge_1992}
{\sc Hoogerbrugge, P.~J \& Koelman, J. M. V.~A} 1992 Simulating microscopic
  hydrodynamic phenomena with dissipative particle dynamics. {\em Europhysics
  Letters ({EPL})\/} {\bf 19}~(3), 155--160.

\bibitem[Jeffrey(1992)]{jeffrey1992}
{\sc Jeffrey, D.~J.} 1992 The calculation of the low reynolds number resistance
  functions for two unequal spheres. {\em Physics of Fluids A: Fluid
  Dynamics\/} {\bf 4}~(1), 16--29.

\bibitem[Jeffrey \& Onishi(1984)]{jeffrey_onishi_1984}
{\sc Jeffrey, D.~J. \& Onishi, Y.} 1984 Calculation of the resistance and
  mobility functions for two unequal rigid spheres in low-reynolds-number flow.
  {\em Journal of Fluid Mechanics\/} {\bf 139}, 261–290.

\bibitem[Kim \& Karrila(2013)]{kim_karrila}
{\sc Kim, Sangtae \& Karrila, Seppo~J} 2013 {\em Microhydrodynamics: principles
  and selected applications\/}. Courier Corporation.

\bibitem[Lees \& Edwards(1972)]{Lees_Edwards_1972}
{\sc Lees, A~W \& Edwards, S~F} 1972 The computer study of transport processes
  under extreme conditions. {\em Journal of Physics C: Solid State Physics\/}
  {\bf 5}~(15), 1921--1928.

\bibitem[Luding(2008)]{Luding2008}
{\sc Luding, Stefan} 2008 Cohesive, frictional powders: contact models for
  tension. {\em Granular Matter\/} {\bf 10}~(4), 235.

\bibitem[Mari {\em et~al.\/}(2014)Mari, Seto, Morris \&
  Denn]{Mari_Seto_2014JoR}
{\sc Mari, Romain, Seto, Ryohei, Morris, Jeffrey~F. \& Denn, Morton~M.} 2014
  Shear thickening, frictionless and frictional rheologies in non-brownian
  suspensions. {\em Journal of Rheology\/} {\bf 58}~(6), 1693--1724.

\bibitem[Marschall {\em et~al.\/}(2019)Marschall, Keta, Olsson \&
  Teitel]{Marschall_etal_2019}
{\sc Marschall, Theodore, Keta, Yann-Edwin, Olsson, Peter \& Teitel, S.} 2019
  Orientational ordering in athermally sheared, aspherical, frictionless
  particles. {\em Phys. Rev. Lett.\/} {\bf 122}, 188002.

\bibitem[Mewis \& Wagner(2012)]{mewis_wagner_book}
{\sc Mewis, J \& Wagner, N~J} 2012 {\em Colloidal suspension rheology\/}.
  {C}ambridge {U}niversity {P}ress.

\bibitem[Morris(2020)]{Morris_annurev2020}
{\sc Morris, Jeffrey~F.} 2020 Shear thickening of concentrated suspensions:
  Recent developments and relation to other phenomena. {\em Annual Review of
  Fluid Mechanics\/} {\bf 52}~(1), null.

\bibitem[Ness {\em et~al.\/}(2018)Ness, Mari \& Cates]{Ness_Mari_Cates_2018}
{\sc Ness, Christopher, Mari, Romain \& Cates, Michael~E.} 2018 Shaken and
  stirred: Random organization reduces viscosity and dissipation in granular
  suspensions. {\em Science Advances\/} {\bf 4}~(3).

\bibitem[Radhakrishnan(2018)]{ranga}
{\sc Radhakrishnan, Rangarajan} 2018 rangrisme/lubrication: Lubrication force.

\bibitem[Royer {\em et~al.\/}(2015)Royer, Burton, Blair \&
  Hudson]{Royer_etal_2015}
{\sc Royer, John~R., Burton, George~L., Blair, Daniel~L. \& Hudson, Steven~D.}
  2015 Rheology and dynamics of colloidal superballs. {\em Soft Matter\/} {\bf
  11}, 5656--5665.

\bibitem[Seto {\em et~al.\/}(2013)Seto, Mari, Morris \& Denn]{Seto_PRL2013}
{\sc Seto, Ryohei, Mari, Romain, Morris, Jeffrey~F. \& Denn, Morton~M.} 2013
  Discontinuous shear thickening of frictional hard-sphere suspensions. {\em
  Phys. Rev. Lett.\/} {\bf 111}, 218301.

\bibitem[Singh {\em et~al.\/}(2019)Singh, Pednekar, Chun, Denn \&
  Morris]{Singh_attr_prl2019}
{\sc Singh, Abhinendra, Pednekar, Sidhant, Chun, Jaehun, Denn, Morton~M. \&
  Morris, Jeffrey~F.} 2019 From yielding to shear jamming in a cohesive
  frictional suspension. {\em Phys. Rev. Lett.\/} {\bf 122}, 098004.

\bibitem[Trulsson(2018)]{trulsson_2018}
{\sc Trulsson, Martin} 2018 Rheology and shear jamming of frictional ellipses.
  {\em Journal of Fluid Mechanics\/} {\bf 849}, 718–740.

\bibitem[Verlet(1967)]{Verlet1967}
{\sc Verlet, Loup} 1967 Computer ``experiments'' on classical fluids. {I}.
  {T}hermodynamical properties of {L}ennard-{J}ones molecules. {\em Phys.
  Rev.\/} {\bf 159}, 98--103.

\end{thebibliography}

\end{document}